# An Integrated Model for User Innovation Knowledge Based on Super-network

Xiao Liao[1], Zhihong Li[1], Yunjiang Xi[1], Haibo Wang[2], Kenneth Zantow*[3]

[1] School of Business Administration, South China University of Technology, 510641 GZ China

[2] A.R.Sanchez, Jr. School of Business, Texas A&M International University, 78041 TX USA

[3] College of Business, University of Southern Mississippi, 39560 MS USA

**Abstract:** Online user innovation communities are becoming a promising source of user innovation knowledge and creative users. With the purpose of identifying valuable innovation knowledge and users, this study constructs an integrated super-network model, i.e., User Innovation Knowledge Super-Network (UIKSN), to integrate fragmented knowledge, knowledge fields, users and posts in an online community knowledge system. Based on the UIKSN, the core innovation knowledge, core innovation knowledge fields, core creative users, and the knowledge structure of individual users were identified specifically. The findings help capture the innovation trends of products, popular innovations and creative users, and makes contributions on mining, and integrating and analyzing innovation knowledge in community based innovation theory.

## 1. Introduction

Users are highly innovative and often innovate or create new products because the users are more conscious about their needs. Nowadays, more and more companies are trying to follow the trend to adopt users and their innovations in their new product development (NPD). As the internet facilitates the interaction between users and the production of user generated content (UGC), opinions and innovations created by users about products are being shaped in real time amongst users through online communities.

As many scholars have indicated, user innovation, which is also defined as customer innovation, open innovation and co-creation, plays a significant role in the process of product innovation and NPD. In terms of online community based user innovation, prior research has provided abundant findings on the co-creation value of users and on how to motivate user innovation and user interaction in online communities. Online user communities are becoming promising sources of user innovations [1-3] which leads to the following research question:" What are the innovations and who are the innovators?" Recent studies mainly emphasize capturing the innovative ideas [4-6] and interactions among users in communities [7-9].

As more and more innovative users gathered in online innovation communities, users began to share what they have developed or innovated through their postings rather than just providing innovative ideas. Some communities even have a special space for these users to submit and share their innovations. However, many firms do not have very clear criteria to judge the innovations and suffer from a lack of systematic standards to evaluate them. For example, what are the main focuses and fields of the innovations? What are the innovative posts and who are the creative innovators related to the hot innovations? While the prior research on online user innovation communities has mainly focused on identifying the motivation of users' innovation

*Corresponding Author:

Email:winnie3223@163.com; ken.zantow@usm.edu

[10-11], there is limited research identifying and integrating user innovation knowledge in a systematic view.

Therefore, our study attempts to construct an integrated model of innovation knowledge in online user innovation communities. Through integrating the users, innovation posts, and innovation knowledge, our study constructs a systematic knowledge integration model that offers a practical tool to identify important or popular innovation knowledge, knowledge fields, the users and posts related to these innovations.

We first identify some online innovation communities as the empirical settings to mine all the posts, users and the relations. We then mine the innovation knowledge keywords from each post. Knowledge mined from posts can be fragmented and lack contextual meaning, which consequently may result in incomplete and fragmented analysis of user innovation activities and decrease the integrity of the measures. To avoid this fragmentation, our research will use the super-network model as an integrated tool.

Super-network was defined as "the networks above and beyond the existing networks" by Nagurney [12]. It can be used to describe and analyze the structure of systems composed of multi-type or multi-tie networks. In online innovation communities, the relations among users, innovation posts, innovation knowledge and users' interactions frame the innovation knowledge system which is composed of the four elements and their multi-tie relations. Consequently, super-network is not only used to integrate the fragmented knowledge but also construct and analyze the knowledge system in online innovation communities.

The objective of this research is therefore (1) to analyze the different elements and the relations in an online innovation community knowledge system and construct an integrating model of user innovation knowledge, and (2) to construct a novel super-network model by analyzing the characteristic of innovation knowledge in online communities. The research contributes to the community-based innovation theory for mining and integrating user innovation knowledge in a community knowledge system.

## 2. Theoretical Background

### 2.1 User Innovation Communities

The importance of users as a valuable source of innovation has been emphasized from 1980s and then gained much more attention in the past ten years [13-14].

Prior research on user innovation mainly focused on how to motivate users to innovate [15-20] and how to support users as innovators [21-22]. Various ways have been used to invite users during the innovation process, such as providing them with toolkits [23]or virtual customer environments [24], talking to lead users [25], and establishing online user communities [26]. Recently, with more and more innovative users gathered in online communities, user innovation in online communities has been receiving more and more attention from researchers [10, 27-30].

Online user innovation communities are groups of individuals who share similar interests and needs and be seen as valuable sources of innovation because of their product experience and cooperation [31]. The innovation behavior of communities members is influenced by their innovation willingness, perceived control and group norm factors [32](Zhang, 2012), and highly motivated by the prospect of improvements in their use of the product [33]. To be seen as a hobbyist and be recognized by the firm are two important motivations community. Members tend to reveal and share their ideas and innovations in communities, join the firm's domain, and engage in innovation process [34]. Community members intend to gain high reputations in the eyes of others such as colleagues and other community members or related companies, so the chance to exchange ideas and experiences and to gain reputation are also the main motivations for community members [35-38].

Dong & Wu（2015）analyzed two online user innovation community-enabled capabilities: ideation capability in collecting user-generated ideas about potential innovation from online user innovation communities, and implementation capability in selecting user-generated ideas for innovation development and introducing developed innovation via online user innovation communities. They found that ideation capability actually does not influence firm value, whereas implementation capability increases firm value [39].

Since online user innovation communities have been drawing an increasing number of attention from academic scholars and firms, these communities have become a vital source of product and service innovations and have helped innovation-related knowledge and innovative users to blossom[40-42]. Therefore, an approach to integrate the innovation knowledge and innovative users in order to improve the firm's implementation capability of online user innovations is needed.

**2.2 Knowledge Super-network**

Nagurney and Dong(2002) defined super-network as being "above and beyond" the existing network, and can be used to solve the decision problem in multi intertwined networks[12]. Other researchers regarded networks which can be described as hyper-graphs as a super-network.

Since the concept of the super-network can describe the structure of complex network system more accurately, and reflect the interaction and impact between the networks, it provides a new way and useful tool to study complex network system and gain gradual attentions from researchers. Knowledge super-network research mainly focuses on the study of network properties and the inter-relations between networks in knowledge spread, share and transfer management.

In the prior research, Knowledge Super-Network(KSN) models are comprised of three types of nodes and six types of edges, reflecting knowledge, persons, material storage media, and the complex interrelations between them. The KSN model can describe complex compositions and the structure of organizational knowledge systems [43]. Yu (2008) constructed a super-network model of organizational personnel training, in which there were three types of networks: the organization personnel network, the material carrier network and the knowledge network [44]. Liao et.al (2011) purposed a super-network model of organizational knowledge sharing network based on a knowledge network and a social network [45]. Ni et.al (2013) constructed a super-network model of Wiki ontology based on the concepts and their multi relations [46]. Wang (2015) presented an improved knowledge diffusion hyper-network (IKDH) model and found out that the diffusion speed of IKDH model is 3.64 times faster than that of traditional knowledge diffusion (TKDH) model [47].

Based on the prior researches, we present an improved super-network model of online user innovation communities (UIWKSN) based on the community's interactions and the characteristics of user innovations.

**3. The Super-network Model of User Innovation Knowledge System**

In the opinion of system science, users' innovation activities can be viewed as an innovation knowledge system which is composed of four elements: users, posts, innovation knowledge and innovation knowledge fields. The elements and their relations can be expressed as fig.1:

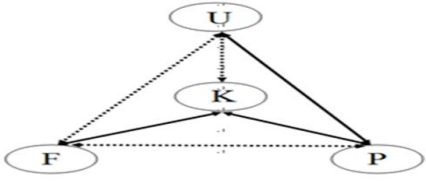

Fig. 1 Composition of UI Knowledge System

In fig.1, it can be seen that the UI knowledge system is composed of 4 kinds of elements and 6 relations, which can be described as a super-network model:

$$UIKSN = (V, E, Q) \quad (1)$$

The expression of each element and relation is defined and explained as:

$V = (U, P, K, F)$ denotes different types of vertices, in which $U = \{u_1, u_2, \cdots, u_n\}$ denotes the set of community users, $P = \{p_1, p_2, \cdots, p_m\}$ denotes the set of posts; $K = \{k_1, k_2, \cdots, k_c\}$ denotes the set of innovation knowledge points mining from posts; $F = \{f_1, f_2, \cdots, f_b\}$ denotes the set of innovation knowledge fields integrated from knowledge points above.

$E = (E_{u-p}, E_{u-f}, E_{u-k}, E_{p-k}, E_{p-f}, E_{k-f}, E_{f-f}, E_{k-k})$ denotes different types of edges, in which $E_{u-p} = \{(u_i, p_j) | \alpha(u_i, p_j) = 1\}$ denotes the relation set of users and posts, in which $\alpha(u_i, p_j) = 1$ means user $u_i$ is the author of post $p_j$; $E_{p-k} = \{(p_i, k_j) | \beta(p_i, k_j) = 1\}$ denotes the relation set of posts and innovation knowledge points, in which $\beta(p_i, k_j) = 1$ means the post $p_i$ contains the knowledge point $k_j$; $E_{p-f} = \{(p_i, f_j) | \chi(p_i, f_j) = 1\}$ denotes the relation set of posts and innovation knowledge fields, in which $\chi(p_i, f_j) = 1$ means that the post $p_i$ is related to the innovation knowledge field $f_j$; $E_{k-f} = \{(k_i, f_j) | \delta(k_i, f_j) = 1\}$ denotes the relation set of knowledge points and knowledge fields, in which $\delta(k_i, f_j) = 1$ means the knowledge point $k_i$ belongs to the knowledge field $f_j$; $E_{k-k} = \{(k_i, k_j) | \varepsilon(k_i, k_j) = 1\}$ denotes the relation set of knowledge points, in which $\varepsilon(k_i, k_j) = 1$ means that the knowledge point $k_i$ and $k_j$ co-exist in the same posts; $E_{f-f} = \{(f_i, f_j) | \phi(f_i, f_j) = 1\}$ denotes the affiliation relation set of innovation fields, in which $\phi(f_i, f_j) = 1$ means the field $f_i$ is the direct subordinate to field $f_j$.

$Q = (Q(K), Q(U), Q(F), W(E_{k-k}))$ denotes the weights of vertices and edges, in which $Q(K) = \{q(k_i)\}$ denotes the set of knowledge points' weight, in which $q(k_i)$ denotes the weight of the knowledge point $k_i$; $Q(U) = \{q(u_i)\}$ denotes the set of knowledge fields' weight, $u_i$ denotes the weight of knowledge field $f_i$; $Q(F) = \{q(f_i)\}$ denotes the set of knowledge fields' weight, $q(f_i)$ denotes the weight of knowledge field $f_i$; $W(E_{k-k}) = \{w(k_m, k_n) | m, n = 1, \cdots, l\}$ denotes the set of links' weights of knowledge points.

Therefore the UIKSN model can be expressed as:

$$\begin{aligned}UIKSN &= (V, E, Q) \\ &= (U, P, K, F, E_{u-p}, E_{p-k}, E_{k-f}, E_{f-f} \\ &\quad Q(K), Q(U), Q(F), W(E_{k-k}))\end{aligned} \quad (2)$$

## 4. The Construction of UIKSN

The construction of UIWKSN includes 4 steps:

(1) Data collection: we use the software of LocoySpider to collect the data including users and their innovation posts in the selected user innovation community. The data of innovation posts is composed of the post titles, content, authors, posting time, reviewing and replying numbers.

(2) Content mining: first, we encode authors and posts to get the user set of U and posts set of P; second, we identify word segments for each post to get keywords to develop the knowledge set of K after preprocessing the keywords; third, we calculate the frequency of each keyword and get the set of knowledge weights Q.

(3) Construction of knowledge network: we construct a weighted knowledge network (WKN) based on the knowledge set, the user attention degree set and the co-occurrence relations between knowledge points; then based on the WKN model, knowledge is clustered into several groups (each group represents a knowledge field) that weighted knowledge clustering network (WCN) is constructed.

(4) Construction of super-network: we calculate the relations between different elements and construct the weighted knowledge super-network of user innovation community (UIKSN).

5. The Analysis of UIKSN in Xiaomi Community

5.1 Data Mining in Xiaomi community

Our empirical setting is the Xiaomi community which was created by the Xiaomi company aiming to provide a platform where the brand's most hardcore users, dubbed "Mi fans," meet to discuss gadgets, and share knowledge. The Xiaomi community is a famous site of user innovation communication in China. In the community, there is a special section that allows users to present their innovations for products. We use the software LocoySpider 8.2 to collect all the posts released before April 29, 2015. After preprocessing of all the posts, a total of 16741 posts were found and we determined that 5715 of those posts to be excellent posts.

Considering the quality of posts, we use excellent posts and 30% of the non-excellent posts in our research. In Xiaomi Community, the excellent posts are identified and marked by their Community staffs. The total number of the posts used in this research is 9024, with 3824 users in them.

First, the posts set (P) and the users set (U) were encoded as $P = \{p_1, p_2, ......, p_{9024}\}$ and $U = \{u_1, u_2, ......u_{3824}\}$.

Second, the software of NLPIR was used to segment the content of each post and get at least 5 keywords for each post. Then we got the knowledge set (K) and their frequency set (Q(K)). According to the selection rule of high frequency words, we totally got 47 high frequency words as the hot keywords, with the lowest frequency of 200. The hot keywords is as table 2.

5.2 Construction of UIKSN

The construction of UIWKSN is based

Table 2. Hot Keywords and Their Frequency in Xiaomi Community

| Knowledge points | Freq. | Knowledge points | Freq. | Knowledge points | Freq. | Knowledge points | Freq. | Knowledge points | Freq. |
|---|---|---|---|---|---|---|---|---|---|
| Function | 1264 | Method | 476 | Support | 363 | Input | 248 | Data Volume | 223 |
| System | 1154 | Desktop | 468 | Manage | 356 | Intelligent | 244 | Mobile | 219 |
| Files | 891 | Interface | 459 | Picture | 329 | Phone | 243 | Router | 217 |
| Set up | 780 | Program | 455 | Icon | 328 | Operate | 241 | Modify | 217 |
| Downloa | 708 | Games | 437 | Time | 316 | Music | 233 | Package | 212 |
| Select | 707 | Screen | 430 | Display | 299 | Headset | 231 | Service | 212 |
| App | 658 | Data | 423 | Users | 282 | Password | 231 | Information | 208 |
| Computer | 643 | Themes | 413 | Test | 281 | Memory | 230 | | |
| Links | 564 | Battery | 399 | Video | 280 | Version | 227 | | |
| Networks | 506 | Mode | 368 | Effect | 250 | Share | 225 | | |

on the construction of the WKN model in knowledge level and the WCN model in knowledge field level. After integrating all the relations between each element and the WKN and WCN model, the UIWKSN in Xiaomi community can be constructed as followed:

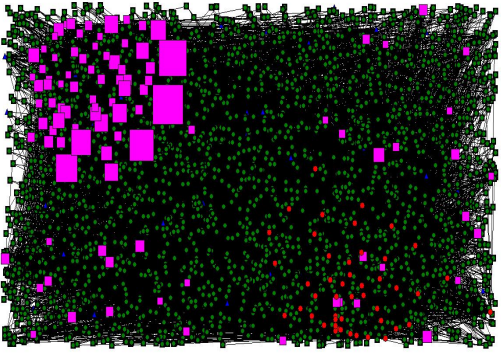

Fig 2. UIWKSN in Xiaomi Community (Part)

In figure 2, the red circle nodes represent knowledge points; the purple rectangle nodes represent users; the green square nodes represent posts, and the blue triangle nodes represent knowledge fields.

## 6. The Analysis of Knowledge Integration Based on UIKSN

### 6.1 The Discovery and Integration of UI Hot Knowledge

UI Hot Knowledge means the innovation knowledge that users create most in community. In WKN, it can be shown as the knowledge points with higher frequency.

$$K_1 = \{k_i \mid q(k_i) \geq q_0, i = 1, \cdots, n\} \quad (3)$$

The users and posts related to the hot knowledge $k_i$ can be expressed as:

$$P(k_i) = \{p_a \mid \alpha(p_a, k_i) = 1, k_i \in K_1\} \quad (4)$$
$$U(k_i) = \{u_b \mid \beta(u_b, k_i) = 1, k_i \in K_1\} \quad (5)$$

Then, according to the relations between knowledge points and the knowledge fields, we can also figure out the hot knowledge related fields.

Let $L(k_i, f_0)$ denote the path from the knowledge $k_i$ to the cluster $f_0$, and then $FL(k_i, f_0)$ represent the set of nodes in these paths and $EL(k_i, f_0)$ represent the set of links between knowledge point $k_i$ and field $f_0$. Then $L(k_i, f_0)$ can be expressed as:

$$L(k_i, f_0) = (FL(k_i, f_0), EL(k_i, f_0)) \quad (6)$$

$L(k_i, f_0)$ clearly show the subordinate relations between hot knowledge $k_i$ and the knowledge field it belongs to.

The analysis of hot innovation knowledge $k_i$ can also be done through the construction of ego super-network.

$$KSN(k_i) = \{V(k_i), EV(k_i), QV(k_i)\} \quad (7)$$

Here, $V(k_i) = (P(k_i), U(k_i), FL(k_i))$ denotes the nodes set of users, posts and fields that related to knowledge $k_i$; $EV(k_i)$ denotes the set of links between the nodes in $V(k_i)$ and $QV(k_i)$ denotes the set of weights of nodes and links in $V(k_i)$.

Based on the above analysis method, we chose 47 core knowledge points (see Table 2) as the hot innovation knowledge to construct the hot knowledge super-network in Xiaomi community was as followed:

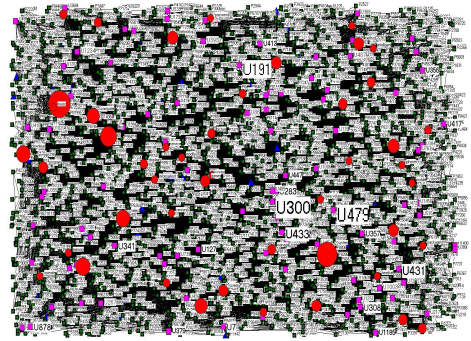

Fig 3. The Hot Knowledge sub-KSN in Xiaomi Community (Part)

There are 3067 related users and 2816 posts with these 47 hot knowledge points. In this sub-graph, the size of the node's label represents the total contribution of each user in these 47 hot knowledge categories. So we can clearly figure out that user 479 and user 300 made the greatest contribution in these hot knowledge, followed by user 191, user 431 and user 433. Through the super-network, we can also find out the hot knowledge related posts of each user. If we want to know more about a single user or knowledge point, we can construct the ego super-network to have a deep analysis. Figure 4 is a ego super-network of the node "screen".

The ego super-network was constructed by calculating the shortest path between the node "screen" and the related knowledge points, users, posts, and knowledge field. From figure 4, we can see that user 366 and user 7 made the greatest contribution in screen innovation, followed by user 191, user 269, user 479, user 535, user 413 and user 341. These users and their posts can be taken into consideration when the company is going to innovate the mobile's screen later

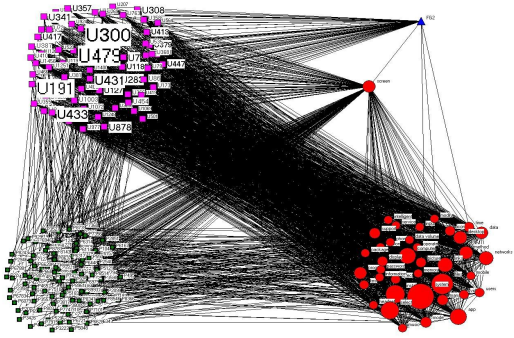

Fig 4. Ego Super-network of "Screen"

### 6.2 The Discovery of Core Innovative Users of Community

The core innovative users in community are the users who create and make great contribution to the core innovation knowledge in community. Through the UIWKSN, the core innovative users can be identified.

First, we need to find out the core innovation knowledge which can be expressed as a sub-WKSN:

$$WKN' = \{K', W(K'), E(K'), W(E')\} \quad (8)$$

$K'$ denotes the set of core innovation knowledge and $W(K')$ denotes the set of nodes' weights.

$$K' = \{k_j \mid q(k_j) \geq q_0\} \quad (9)$$

Here, $q_0$ is the threshold value which is used to judge the core innovation knowledge.

$E(K')$ denotes the set of links between core innovation knowledge and $W(E')$ denotes the set of links' weights.

$$E(K') = \{(k_i, k_j) \mid k_i \in K', k_j \in K'\} \quad (10)$$

Based on the relations among WKN', users and posts, we can construct the sub-WKSN of core innovation knowledge super-network in the community. Then, the related users and posts of core innovation knowledge can be identified easily.

$$WKSN(K') = \{K', P(K'), U(K'), E(K'), E_{p-p}(K'), E_{u-u}(K'), E_{p-k}(K'), E_{u-p}(K'), E_{u-k}(K'), W(K'), W(E_{p-k}(K')), W(E_{u-k}(K'))\} \quad (10)$$

The set of core knowledge related users is expressed as:

$$U(K') = \{u_i \mid k_j \in K', \phi(u_i, q') = 1, q(u_i, k_j) \geq q'\} \quad (11)$$

Here, $k_j$ represent the core knowledge points; $q(u_i, k_j)$ represent the related level of user $u_i$ and $k_j$; $q'$ is a threshold value which used to judge the core knowledge related users.

The set of $K'$ related posts is:

$$P(K') = \{p_b \mid k_j \in K', \gamma(p_b, k_j) = 1, q(p_b, k_j) \geq q_1\} \quad (12)$$

Here, $q_1$ is also a threshold value which is used to judge the core knowledge related posts.

Based on the relations between the 47 core knowledge points and the related users and posts, the sub-WKSN can be constructed. Then, by setting the threshold value of the relations between core knowledge and users, we can clearly figure out the core innovative users in community (see fig. 5)

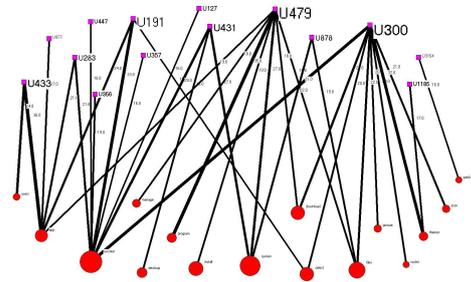

Fig. 5 Sub-WKSN of Core Innovative Users
（edges' weight higher than 15）

In figure 5, the size of knowledge points represents the knowledge frequency in posts; the size of users' labels represents the total contribution in these 47 core innovation knowledge points of each user; the size of edge between users and core knowledge points represents the users' contribution in each core knowledge point. Thus, we can clearly find out the main core users in the community. For example, knowledge "function" and "system" have the similar word frequency, but the number of related users is far different. There are many more related users on the innovation of "system" than "function". Although there are few users on the innovation of "function", there are more users who create innovations on it more than once. Next, the frequency of the knowledge of "application" is about half of the frequency of "system", but the number of related core users who create innovations on it more than once is much more than "system", which implies that only some core users concentrate on the innovation of "application" and innovate it repeatedly.

Furthermore, from this figure, we can also identify the users' main innovation knowledge. For example, users "479 " and "300" made the most contributions on the core knowledge, while their innovation focuses are different. User " 479" mainly focuses on the innovation of "program" and "system"; while user "300" mainly focuses on "function" and "theme".

Therefore, through the sub-WKSN of the core knowledge, we can not only identify the core innovative users, but also identify their main focus on innovations.

### 6.3 The Discovery of Hot Innovation Fields of Community

The hot innovation fields can be constructed as the sub-WCN with higher cluster weights which is expressed as:

$$F_1 = \{f_i \mid q(f_i) \geq q_2, i = 1, \cdots, c\} \quad (13)$$

$FV(f_i)$ denotes the set of knowledge fields and their knowledge points included in the hot innovation field $f_i$. $K_2(f_i)$ denotes the knowledge points directly or indirectly included in $f_i$, which can be expressed as:

$$K_2(f_i) = \{k_a \mid \gamma(k_a, f_i) = 1\} \quad (14)$$

Based on the UIKSN, the hot innovation fields related users and posts can also be identified:

The posts related to hot fields $f_i$:
$$P_2(f_i) = \{p_j \mid \alpha(p_j, k_a) * \gamma(k_a, f_i) = 1\} \quad (15)$$

The related users:
$$U_2(f_i) = \{u_j \mid \alpha(u_c, p_j) * \beta(p_j, k_a) * \gamma(k_a, f_i) = 1\} \quad (16)$$

In order to analyze the hot innovation fields, it can also be constructed to a sub-WKSN:

$$KSN(f_i) = \{V(f_i), EV(k_i), QV(f_i)\} \quad (17)$$

Here, $V(f_i) = (P(f_i), U(f_i), K(f_i))$ denotes the nodes' set of related users, posts, fields to the field $f_i$; $EV(f_i)$ denotes the set of relations among nodes in $V(f_i)$; $QV(f_i)$ denotes the set of nodes and edges weights in $V(f_i)$.

The sub-KSN can integrate all the related essences to the hot innovation fields that we can find out most of the useful information in these hot fields. In the UIWKSN of Xiaomi community, there are 8 innovation fields with 104 sub-fields in them. Here, we chose the fields "FA251" and "FA6" as the two hot innovation fields with the threshold of $q_2 \geq 10000$. In these two hot fields, there are 3749 related users and 8931 related posts. Through the calculation of all these users' contribution on the two fields, we can identify the core innovative users in the hot fields. Figure 6 is the sub-KSN of hot innovation fields in Xiaomi community with 90 innovative users whose contributions are the head 30% of the whole related users.

In figure 6, the size of the nodes' label represents the word frequency or the total contributions in the hot fields of each user. Therefore, we can clearly see that the core innovative users who made great contribution in hot fields include users "U479", "U4 31", "U433", "U447", "U878" and so on.

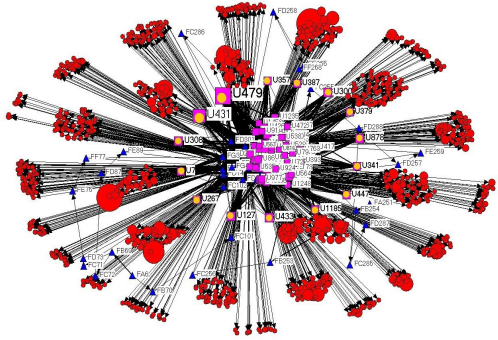

Fig. 6 The Sub-KSN of Hot Innovation Fields in Xiaomi Community

Based on the sub-KSN of hot innovation fields, we can have a better understanding on the relations between the hot innovation fields and the related core innovative users through calculating the edges' weights between them, which implies the contributions each user made in the related fields. Figure 7 is the sub-KSN of hot innovation fields and the related users in which the threshold of edges' weight is 20.

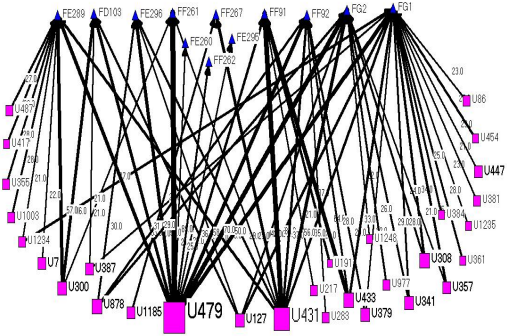

Fig. 7 The Sub-WKSN of Hot Innovation Fields and the Related Users (edges' weight higher than 20)

In figure 7, we can see that user "U479" is the core innovative users in hot fields who made the most contributions in most of the hot fields such as "FG1", "FG2", "FF261" and "FE289". And among the hot fields, "FG1" has the most related users followed by "FE289" and "FG2". While the fields "FF261" and "FF91" have fewer related users but some of the users made repeat contributions on them such as users "U479", "U431", "U433" and "U300". Thus these users can be seen as the core innovative users in these two fields.

### 6.4 The Integration and Analysis of the Users' Self Innovation Knowledge

#### 6.4.1 The User's Self UIKSN Model

The online communities gather a large amount of users who have different expertise. Through the UIKSN, we can construct the sub-KSN model for each user so that we can know more about his or her innovation knowledge structure. Based on the relations between U-F, U-P and U-K, the sub-WKSN of each user can be expressed as:

$$WKSN(u_i) = \{u_i, F(u_i), K(u_i), E_{k-k}(u_i), E_{u-f}(u_i), E_{f-k}(u_i), E_{u-k}(u_i), W(K(u_i)), W(F(u_i)), W(E_{k-k}(u_i)), W(E_{u-k}(u_i))\} \quad (18)$$

This sub-KSN is a super-network including the 4 elements (user, posts, innovation fields and innovation knowledge) in community's knowledge system which is helpful in the identification of user's knowledge structure and expertise when companies need to invite users to their innovation or NPD process.

#### 6.4.2 The Construction of User's Self UIKSN Model

Based on the analysis of the above hot innovation knowledge and hot innovation fields, we found that user "U479" is one of the users who made the most contributions. Thus, we chose "U479" and the other randomly chosen user "U127" as the two subjects to construct their UIKSN model. Table 2 lists the two users' number of related posts, knowledge points and the innovation fields.

Table 2 The Innovation Info of "U479" and "U127"

Based on the basic information in table 2, the two users' UIKSN was constructed and shown

| User | No. of P | No. of K | No. of F |
|---|---|---|---|
| U127 | 46 | 127 | 32 |
| U479 | 143 | 272 | 43 |

in figure 8 and 9.

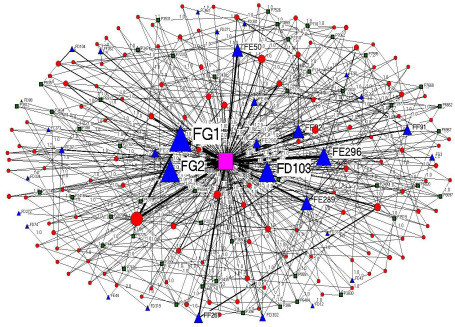

Fig. 8 The Self UIKSN of "U127"

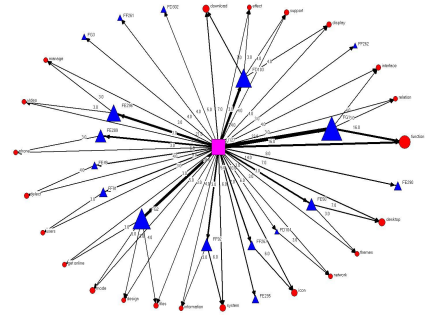

Fig. 10 The Core Innovation Knowledge Super-network of "U127"

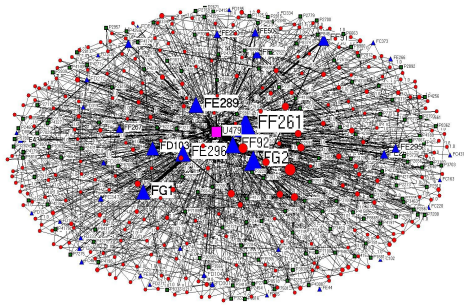

Fig. 9 The Self UIKSN of "U479"

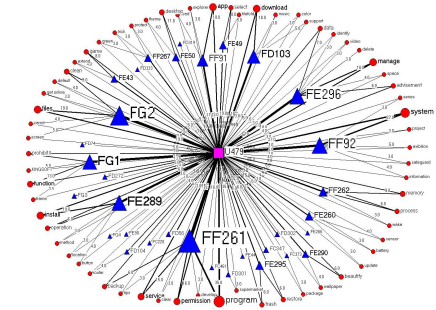

Fig. 11 The Core Innovation Knowledge Super-network of "U479"

Since the user "U479" had more posts, the self UIKSN is much more complicated than "U127". From the figures, we know that the main innovation fields of "U127" focused on "FG1", "FG2", "FE296" and "FD103" with the most contributions on knowledge "function", "system", "desktop", "icon" and "download"; while the main innovation fields of "U479" focused on "FF261", "FG2", "FF92", "FG1", "FE289" and "FE296" with the most contributions on knowledge "program", "application", "system", "permission", "files", "download" and "management".

Through the construction of user's sub-UIKSN, we can have more specific analysis such as the user's core innovation knowledge, core innovation field.

**6.4.3 The Identification of User's Self Core Innovation Knowledge**

Based on the basic rule of high frequency words selection, the super-network of each user's core knowledge was constructed and drew as figure 10 and 11.

Figure 10 and 11 clearly show us the core innovation knowledge and core innovation fields of these two users. For example, the core innovation fields of "U127" are "FG1", "FG2", "FE296" and the related core innovation knowledge are "function", "interface", "files", "design", "mode", "video" and so on; while the core innovation fields of "U479" are "FF261", "FG2", "FG1", "FE296" and the related core innovation knowledge are "program", "backup", "permission", "screen", "files", "clear", "space", "advertisement", "data" and so on.

**6.4.4 The Analysis of Ego UIKSN**

Besides the user's self UIKSN model, we can also construct the ego UIKSN model for each innovation knowledge, innovation field and innovation post to have a deep analysis. For example, the ego super-network of a user in a certain innovation field (see fig. 12-13); the ego super-network of a user in a certain innovation knowledge and the related posts (see fig. 14-15); the ego super-network of a user in a certain post and the related knowledge (see fig. 16-17).

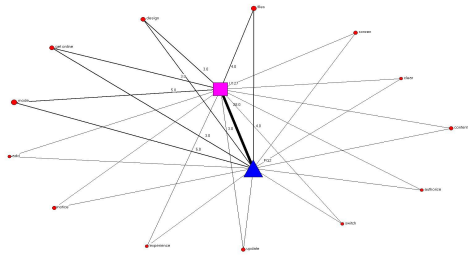

Fig. 12 The Ego UIKSN of "U127" in "FG2"

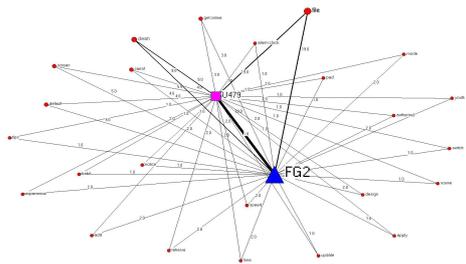

Fig. 13 The Ego UIKSN of "U479" in "FG2"

From the analysis above, we know "FG2" is the core innovation field of users "U127" and "U479". But the figures 15-16 showed that the two users have different innovation focus on this field. For example, "U127" mainly focused on "files", "design" and "mode"; while "U479" mainly focused on "files", "clear", "screen" and "internet access".

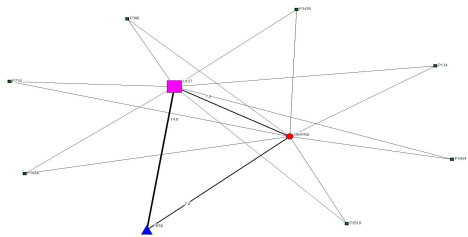

Fig. 14 The Ego UIKSN of "U127" in "Desktop"

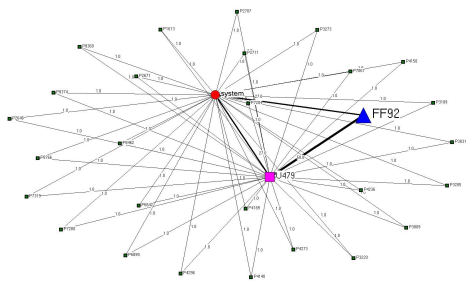

Fig. 15 The Ego UIKSN of "U479" in "System"

Figures 14-15 showed us the two users' innovations on a certain knowledge such as "desktop" and "system". This ego UIWKSN also showed us the knowledge related innovation field and the related posts. So if we want to know more details in this knowledge area, we can find the related posts.

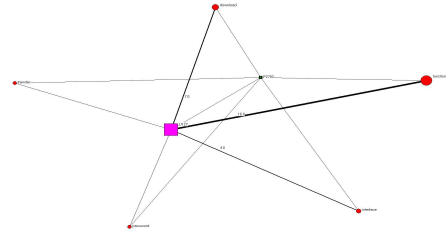

Fig. 16 The Ego UIKSN of "U127" in "P2782"

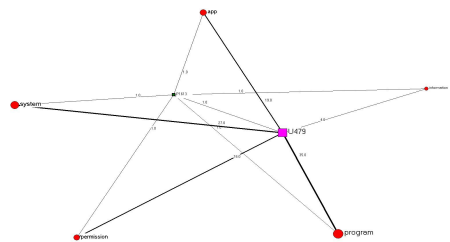

Fig. 17 The Ego UIKSN of "U479" in "P1613"

Figure 16-17 showed us the two users' innovations in a certain post. In this ego UIKSN, we can easily identify the main knowledge in the post. For example, the innovation knowledge in "P2782" posted by "U127" are "function", "desktop", "download", "password" and "delivery"; while the innovation knowledge in "P1613" posted by "U479" are "system", "program", "application", "permission" and "information".

## 7. Conclusion

This paper aims to provide an knowledge integrating method in online user innovation communities by constructing the UIWKSN model. It contributes to the existing literature on user innovation in two ways: (1) by analyzing the elements in communities, we integrate all the elements into a super-network model which provide a new tool for us to integrate the different relations between the elements and (2) by construction the UIKSN, it provides us with a more specific and deeper analysis on the hot and core innovation knowledge, knowledge fields and so on.

The internet as an advanced information and communication medium has led to a large amount of valuable user generated innovation content in the online communities. Literature findings proved that users innovation knowledge and the related innovative users are valuable for the product innovation and the new product development process of companies. Users' innovations not only meet the needs of users but also imply the innovation trend of products. Therefore, how to find out the innovation knowledge created by users is very important for companies. The UIKSN model constructed in this paper helps to integrate and analyze the users' innovation knowledge in these communities. The hot and core innovation knowledge, innovation fields, the core innovative users and the related posts were identified and showed good in the empirical setting: Xiaomi community.

Further research might try to focus on more different communities and consider more elements such as the interaction among users in the knowledge system.

**Acknowledgment:**

This work was supported by the National Natural Science Foundation of China under grant No.71371077 and No.71571073, China Postdoctoral Science Foundation under grant No.2016M602475.